\documentclass[a4paper,conference,10pt]{IEEEtran}
\usepackage{cite}

\usepackage{textcomp}

\usepackage{multirow}

\usepackage[pdftex]{graphicx}

\usepackage{dblfloatfix}

\usepackage[cmex10]{amsmath}

\usepackage{url}

\usepackage[binary-units,per-mode=symbol]{siunitx}
\usepackage{xcolor}
\usepackage{listings}
\definecolor{basiccolor}{rgb}{0, 0, 0}
\definecolor{stringcolor}{rgb}{0.7, 0.1, 0.2}
\definecolor{identifiercolor}{rgb}{0, 0, 0.8}
\definecolor{commentcolor}{rgb}{0.5, 0.5, 0.5}
\definecolor{keywordcolor}{rgb}{0, 0, 0.8}
\definecolor{emphcolor}{rgb}{1, 0, 0}
\lstdefinelanguage{ANDL}{
morekeywords={inline, network, devices, node, switch, gateway, canLink, communication, message, sender, receivers, payload, period, connections, segment, new, mapping, tt, ctID, can, id, pool, holdUp, types, ethernetLink, bandwidth},
sensitive=false,
morecomment=[l]{\%},
morecomment=[l]{//},
morestring=[b]",
identifierstyle=\color{basiccolor},
keywordstyle=\color{keywordcolor}\bfseries,
commentstyle=\color{commentcolor}\itshape,
tabsize=2
}

\lstdefinelanguage{XMLconstraint}{
morekeywords={constraints, constraint, module, name, min, max, avg_min, avg_max, moduleIsRegex, nameIsRegex, samples},
sensitive=false,
morecomment=[l]{\%},
morecomment=[l]{//},
morestring=[b]",
moredelim=[s][\color{keywordcolor}]{>}{<},
identifierstyle=\color{basiccolor},
keywordstyle=\color{keywordcolor}\bfseries,
commentstyle=\color{commentcolor}\itshape,
tabsize=2
}

\lstdefinelanguage{oppini}{
alsoletter={-},
morekeywords={eventlogmanager-class, outputscalarmanager-class, outputvectormanager-class, postgresqloutputmanager-connection},
sensitive=false,
morecomment=[l]{\%},
morecomment=[l]{//},
morestring=[b]",
identifierstyle=\color{basiccolor},
keywordstyle=\color{keywordcolor}\bfseries,
commentstyle=\color{commentcolor}\itshape,
tabsize=2
}

\usepackage{todonotes}

\makeatletter
\def\lst@makecaption{%
  \def\@captype{table}%
  \@makecaption
}
\makeatother

\begin{document}

\title{Extending OMNeT++ Towards a Platform for the Design of Future In-Vehicle Network Architectures}

\author{\IEEEauthorblockN{Till Steinbach, Philipp Meyer, Stefan Buschmann, and Franz Korf}

\IEEEauthorblockA{Department of Computer Science, Hamburg University of Applied Sciences, Germany\\
\{till.steinbach, philipp.meyer, stefan.buschmann, franz.korf\}@haw-hamburg.de}
}

\maketitle

\begin{abstract}
In-vehicle communication technologies are evolving. While today's cars are equipped with fieldbusses to interconnect the various electronic control units, next generation vehicles have timing and bandwidth requirements that exceed the capacities. In particular Advanced Driver Assistance Systems (ADAS) and automated driving using high bandwidth sensors such as cameras, LIDAR or radar will challenge the in-car network. Automotive Ethernet is the most promising candidate to solve the upcoming challenges. But to design and evaluate new protocols, concepts, and architectures suitable analysis tools are required. Especially in the interim period with architectures using automotive Ethernet and legacy fieldbusses together, careful planning and design is of vital importance. Simulation can provide a good understanding of the expectable network metrics in an early development phase.

This paper contributes a workflow as well as the required toolchain to evaluate new real-time Ethernet communication architectures using event based simulation in OMNeT++. We introduce a domain specific language (DSL) -- the Abstract Network Description Language (ANDL) -- to describe and configure the simulation and present the required simulation models for real-time Ethernet and fieldbus technologies such as CAN and FlexRay. We further introduce new analysis tools for special in-vehicle network use-cases and the interaction of the simulation with third-party applications established in the automotive domain.
\end{abstract}

\begin{IEEEkeywords}
Simulation, In-Vehicle Networking, Real-time Ethernet, Automotive Ethernet, CAN, FlexRay, Gateway
\end{IEEEkeywords}

\IEEEpeerreviewmaketitle
\section{Introduction \& Problem Statement}
The in-vehicle network faces a significant paradigm change. While communication architectures of today's vehicles are consisting of different technologies such as Controller Area Network (CAN), FlexRay, Local Interconnect Network (LIN) and Media Oriented Systems Transport (MOST), soon Ethernet will form the backbone for in-vehicle communication. Switched Ethernet is -- due to its high data rate, its low cost of commodity components, and its large flexibility in terms of protocols and topologies -- a promising candidate to overcome the challenges of future in-car networking \cite{mk-ae-15}. But, a sudden change from today's architectures towards a network solely build on Ethernet is impossible with reasonable cost and risk. A consolidation strategy with heterogeneous networks formed of an Ethernet root and legacy busses at the edges will allow to preserve invest in knowledge around legacy technologies. Such a mixed architecture can form the beginning of a stepwise transition from today's bus based designs towards a flat network topology using Ethernet links only.

To design and evaluate such future in-vehicle networks, new tools are required. While current toolchains focus on bit-correct simulation of fieldbus communication, future environments have to enable the developer to analyze effects of congestion and jitter on the cars applications and assistance functions on a system level. The OMNeT++\cite{ah-ooose-08} platform is a well suited tool and a perfect base to implement a flexible workflow. Besides its open-source simulation core, it allows to extend its Eclipse based IDE with custom plugins for specialized design and analysis tasks. With this work we contribute both, a uniform workflow as well as the required models and tools to design and evaluate future in-vehicle networks.

The center of the simulation toolchain are the simulation models that are published open-source. For various real-time Ethernet technologies the CoRE4INET model suite was created. It relies on the OMNeT++ INET framework and provides the implementation of real-time Ethernet protocols as well as clock synchronization. The models for fieldbus technologies are similarly provided in the FiCo4OMNeT suite. To interconnect both technologies -- real-time Ethernet and fieldbusses -- the SignalsAndGateways models were developed.

Experiences with the simulation during research on in-car network architectures showed that the configuration of these large networks is complex and lengthy. Thus there was a demand to simplify the description of in-car network scenarios. This demand led to the development of a domain specific language (DSL) that supports the fast setup of simulations of in-car network architectures.

Finally, in-car networks require the analysis of specific network metrics, for example tracing of jitter in the forwarding chain of cyclic messages. To support the evaluation of in-car networks we created analysis tools and interfaces that support the workflow and allow to pass simulation results to third-party software. While some of these tools are specific to in-car networking, most of the software provided for result analysis is applicable to other network simulation scenarios as well. 

The remaining paper is organized as follows: In Section \ref{sec:background_related}, we introduce the technological background and relate to preliminary and related work. Section \ref{sec:sim_models} presents the simulation models for in-vehicle networks. We present tools for designing and configuring in-vehicle networks in section \ref{sec:network_design} and tools to analyze the simulation results in section \ref{sec:result_analysis}. Using a short case study, section \ref{sec:case_study} presents our workflow. Finally, section \ref{sec:conclusion_outlook} concludes our work and gives an outlook on future research.
\section{Background \& Related Work}
\label{sec:background_related}
Today there are several commercial tools to analyze in-car networks. In industry, most popular is CANoe (by Vector Informatik GmbH) that enables real-time cluster simulations of fieldbusses. Today, CANoe does not provide functionality to simulate real-time Ethernet variants. SymTA/S is a commercial timing analyzer (not a network simulator) by Symtavision GmbH that supports Ethernet (standard and AVB) as well as common fieldbus technologies. It provides analytical models to calculate load and timing.

OMNeT++\cite{ah-ooose-08} is a discrete event based simulation platform mainly focusing on the simulation of networks and multiprocessor systems. It is a perfect base for a simulation tool chain for automotive communication. We developed our model suites as an extension of the popular INET-Framework\cite{inetframework} that provides the implementation of Ethernets physical layer as well as protocols and applications above layer 2.

While there were no publicly available OMNeT++ simulation models for real-time Ethernet technologies, there is another CAN bus model developed independently at the same time at the Nagoya University in Japan. The last release is from April 2014 and is not yet compatible with the latest INET and OMNeT++ releases. The developers also analyzed CAN-Ethernet gateway strategies \cite{kmt-sepea-14}.

For collecting simulation results in databases there are already examples for MySQL provided with OMNeT++. A simple interface for storing results in SQLite is provided with the INET-HNRL, a fork of the INET framework for hybrid networking research \cite{inethnlr}.
\section{Simulation Models}\label{sec:sim_models}
\begin{figure}[!b]
  \centering
    \vspace{-1em}
  \includegraphics[width=\linewidth]{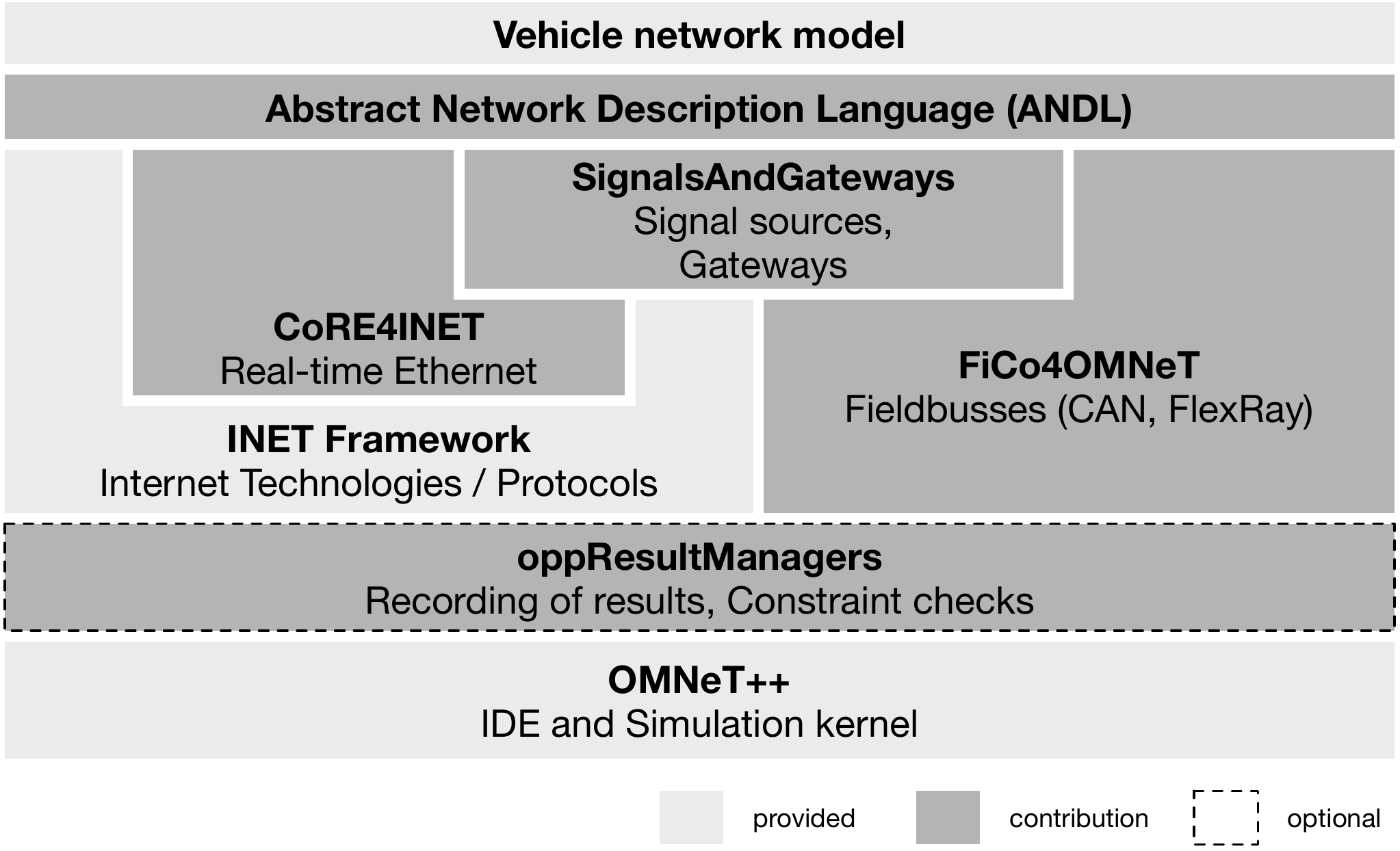}
  \vspace{-1em}
  \caption{Overview of the contributed tools and simulation models}
  \label{fig:models}
\end{figure}

The simulation models introduced in this section were developed for in-car network simulations, but can be used for other systems as well. All models are published open-source (see \url{http://sim.core-rg.de}) and can be used free of charge. Figure \ref{fig:models} gives an overview of the contributed simulation models and their place in the software stack of the toolchain. To simplify the installation an OMNeT++ plugin is provided that offers an automated installation process as well as an update procedure.

\subsection{CoRE4INET}
CoRE4INET (Communication over Real-time Ethernet for INET) is a suite of real-time Ethernet simulation models. Currently it supports the AS6802 protocol suite, traffic shapers of Ethernet AVB, and implementations of IEEE 802.1Q, as well as models to map IP traffic to real-time traffic classes.

The center of the CoRE4INET models is the implementation of media access strategies for different traffic classes. By flexibly combining these strategies, new traffic shapers can be designed that are able to forward real-time traffic of different standards. For example it is possible to combine time-triggered traffic of AS6802 with credit based shaping of Ethernet AVB to form a new time-aware shaper that can handle both classes in parallel \cite{msks-eatts-13}. This allows to evaluate new concepts that are currently under standardization or are even not yet assessed.

For incoming traffic, the models contain traffic selection and constraint checks. To simulate time-triggered behavior and time-synchronization, CoRE4INET provides models for oscillators, timers and schedulers. Oscillators allow to implement the behavior of inaccurate clocks with their unique influence on real-time communication. Finally, CoRE4INET contains application models for simple traffic patterns and traffic bursts.

The simulation models were checked against analytical models of the different specifications and -- where possible -- evaluated in empirical tests using real-world hardware.

CoRE4INET is under active development. New features are constantly added. The next milestone targeting for release soon is the support of frame-preemption as currently discussed in IEEE PAR 802.1Qbu \cite{ieee-802.1qbu}.

\subsection{FiCo4OMNeT}
FiCo4OMNeT (Fieldbus Communication for OMNeT++) is a set of fieldbus simulation models. It was originally developed with separate CAN and FlexRay models, but later merged to take account for the similarities of fieldbus technologies.

Similar to CoRE4INET, FiCo4OMNeT contains models of oscillators and clocks to allow for a precise simulation of the synchronization of FlexRay's static segment. Further it contains application models for CAN and FlexRay applications with simple traffic patterns.
The fieldbus models in FiCo4OMNeT were originally checked against results of the CANoe simulation environment, an industry standard software for the simulation of fieldbus based in-vehicle networks.

Even though fieldbusses are considered legacy technology there are new approaches. With CAN-FD (CAN with flexible data rate) the bandwidth of CAN can be significantly increased. We are currently working on a CAN-FD implementation to enable the simulation of advanced in-car network architectures containing CAN busses with flexible datarate.

\subsection{SignalsAndGateways}
\label{sec:signalsAndGateways}
The SignalsAndGateways simulation models fill the gap between the simulation of real-time Ethernet and fieldbusses. Gateways are nodes that translate between legacy bus technologies and (real-time) Ethernet. To be as flexible as possible, the gateways are divided in three submodules:
\paragraph{Routing} The router module receives messages in their original representation and decides based on forwarding rules which path the message will take. A message can have no routing entry if it should be dropped, one routing entry if it has one receiving bus or node, or several routing entries if it should be visible to several receivers on different busses. There is no limit of busses and links a gateway can be connected to. The gateway can also translate between fieldbus technologies, thus it is also applicable to legacy designs with multiple busses interconnected over a central gateway.
\paragraph{Buffering}
Gateways support aggregation strategies to improve bandwidth utilization of different technologies. CAN messages for example have a maximum payload of \SI{8}{\byte}, while Ethernet messages have a minimum payload of \SI{46}{\byte}. If only one CAN message would be encapsulated in an Ethernet frame, the rest of the frames payload would be padded and bandwidth would be wasted. Aggregation strategies implemented in the buffer modules allow to release frames in groups, according to different strategies. These strategies are implemented in the buffer modules, too.

Aggregation strategies have a huge impact on the latency of messages passing a gateway. All strategies delay frames to collect multiple messages before aggregating them into one large packet. The most popular strategy implemented in the buffer is the pooling strategy with holdup time. Each message is assigned to a pool, while multiple different messages share the same pool. Further each message is assigned a holdup time, representing the maximum acceptable delay for this message. On arrival of a frame in the pool, its holdup time is compared with the pools holdup time. If the frames holdup time is shorter, the pools time is adjusted correspondingly. When the holdup time of the pool is expired all messages in the pool are released together.
The modular architecture of the gateway allows to easily add more aggregation strategies.

\paragraph{Transformation}
Transformation modules implement the translation between different communication technologies. The strategies transparently map information between fieldbusses and Ethernet. Currently there is a simple mapping between fieldbus frames and raw (layer 2) Ethernet frames. The modular architecture of the gateway allows to easily add more sophisticated mappings, e.g. when higher layer application protocols should be used.

Similar to real-world gateways, gateway nodes can host applications that are not related to gateway functionality. Thus gateways can be added to control units that also host application software.
\section{Network Design}
\label{sec:network_design}
Configuring the simulation of large heterogeneous networks is complex and lengthy. To reduce this effort and to let the developer focus on his design task, we developed a domain specific language (DSL) for the description of heterogeneous in-vehicle network designs. It is called \emph{Abstract Network Description Language (ANDL)} and provides an easy and assisted way to design a network in an Eclipse environment. It is implemented as an Eclipse plugin and thus fits into the OMNeT++ IDE. The plugin provides syntax highlighting as well as context aware code completion. For TDMA technologies, the ANDL plugin contains scheduling algorithms that allow to find a first feasible schedule for initial results \cite{ksks-dtina-14}.

Listing \ref{lst:andl} shows an example of a network consisting of two CAN busses interconnected over a real-time Ethernet backbone described in the ANDL.

\begin{lstlisting}[language=ANDL,frame = lines,caption=ANDL code example with comments, label=lst:andl]
types std {           //Types can be defined and reused
  ethernetLink ETH {  //Definition for Ethernet link
    bandwidth 100Mb/s;//Link has bandwidth of 100MBit/s
  }
}   //it is also possible to define types in a separate file

network smallNetwork{ //network name is smallNetwork
  inline ini{         //Inline ini for special parameters
    record-eventlog = false
  }                   //Parameters are inserted into .ini

  devices{            //Define all devices in the network
    canLink bus1;     //First CAN bus
    canLink bus2;     //Second CAN bus
    node node1;       //First CAN node
    node node2;       //Second CAN node
    gateway gw1;      //Gateway for first CAN bus
    gateway gw2;      //Gateway for second CAN bus
    switch switch1;   //Real-time Ethernet Switch
  }
  
  connections{    //Physical connections (Segments = groups)
    segment backbone { //Ethernet Backbone part
      gw1 <--> {new std.ETH} <--> switch1; //Ethernet Link
      gw2 <--> {new std.ETH} <--> switch1; //Ethernet Link
    }
    segment canbus{    //CAN bus part (busses share config)
      node1 <--> bus1; //CAN node connected to first bus
      gw1 <--> bus1;   //Gateway connected to first bus
      node2 <--> bus2; //CAN node connected to second bus
      gw2 <--> bus2;   //Gateway connected to second bus
    }
  }
  
  communication{       //Communication in the network
    message msg1{      //Message definition
      sender node1;    //First CAN node is sender
  	  receivers node2; //Second CAN node is receiver
  	  payload 6B;      //Message payload is 6 Bytes
  	  period 5ms;      //5ms cyclic transmission
  	  mapping{  //mapping to traffic class, id, gw strategy
        canbus: can{id 37;};       //Message ID 37 on CAN
        backbone: tt{ctID 102;};   //TT traffic on backbone
        gw1: pool gw1_1{holdUp 10ms;};   //Aggregation time
  	  }
  	}
  }
}
\end{lstlisting}

\begin{figure}[!t]
  \centering
  \includegraphics[width=\linewidth]{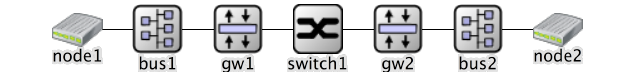}
  \vspace{-2em}
  \caption{ANDL generated network consisting of two CAN busses and a real-time Ethernet backbone with two gateways and one switch}
  \label{fig:smallNetwork}
  \vspace{-1em}
\end{figure}

In comparison to the compact description in ANDL, the size of the generated OMNeT++ config (.ini/.ned/.xml) has more than 250 lines. The resulting network is shown in Figure \ref{fig:smallNetwork}. The definition of the scenario starts with the networks \emph{devices}. Afterwards the previously defined devices are arranged into a network topology in the \emph{connections} section. The topology can be divided in several different segments with different configurations for messages. In the example there is one segment for the Ethernet part called \emph{backbone} and one segment for the CAN bus part called \emph{canbus}. When messages traverse the borders of a segment they are translated from the sending segments representation into the receiving segments representation. The last part of the definition is the actual communication taking place. In the example there is only one message transmitted from \emph{node1} to \emph{node2}. The mapping of each message defines how the message is represented in the different segments. In the example the message is a CAN frame with id \emph{37} on the bus and a time-triggered message with critical traffic id \emph{102} on the real-time Ethernet backbone.

Besides the features shown, the ANDL defines more parameters to describe traffic flows or aggregation strategies. Commonly used components can be defined in include files, e.g. a Ethernet Link with \SI{100}{\mega\bit\per\second}, and used in several places. Further ANDL provides inheritance, thus it is possible to define primitive stencils for components that are later refined during the instantiation.

Currently, ANDL supports only the most commonly used parameters. For more sophisticated configurations inline ini code can be used. Parameters defined in the inline ini sections are directly copied into the resulting omnetpp.ini file.

The ANDL is implemented as an OMNeT++ plugin using Eclipse's Xtext technology. Xtext is a framework for development of programming languages and domain-specific languages. It provides a grammar to define the language and generates the required parsers as well as the code editor for the OMNeT++ IDE.

\section{Result Analysis}
\label{sec:result_analysis}
The OMNeT++ IDE already comes with tools for the result analysis. We extended those built-in tools to simplify the analysis in specialized use-cases and developed interfaces to interconnect the OMNeT++ simulation with established industry products. Our contributions for the result analysis are:
\subsection{Gantt Chart Timing Analyzer}\label{sec:gcta}
The Gantt Chart Timing Analyzer (GCTA) is an OMNeT++ plugin developed to trace jitter and delay in cyclic communication. It uses a timing log file (.tlog) written during the simulation to generate a gantt chart of the communication path from sender to receiver. In contrast to the OMNeT++ eventlog, that shows all events of the simulation in a timeline, the GCTA plugin compresses all occurrences of a cyclic message into one single chart. This allows to easily detect the source of jitter and delay in the path between sender and receiver.

After installing the GCTA plugin in the OMNeT++ IDE, .tlog files (see section \ref{sec:tlog}) recorded during the simulation can be processed. Currently, GCTA only supports analysis of (real-time) Ethernet traffic, but the concept is also applicable to heterogeneous networks with Ethernet and fieldbusses interconnected using gateways. GCTA is an example for a specialized analysis tool implemented as OMNeT++ plugin that extends the built in functionality. As it relies on the visualization capabilities of OMNeT++ no further software is required.

\subsection{oppResultManagers}
\label{sec:oppResultmanagers}
oppResultManagers is a set of modules for OMNeT++ simulations. Instead of simulation models it contains so called ResultManagers. ResultManagers are responsible for writing out simulation results. The OMNeT++ vector and scalar files, as well as the eventlog are built-in instances of ResultManagers. The oppResultManagers project adds the following ResultManagers to OMNeT++:
\subsubsection{PCAPng} The PCAP next generation  (PCAPng) dump file format \cite{pcapng-draft} is an attempt to overcome the limitations of the currently widely used (but limited) libpcap format. Libpcap allows to log packet oriented communication and is used in popular analysis tools such as Wireshark. The most important extension of PCAPng is the support of multiple interfaces in one file. The INET framework already contains a module to write legacy PCAP files, but it only supports communication above IP layer. Due to the PCAPng module being a ResultManager, no changes to the simulation must be made to write PCAPng files, it is simply enabled in the ini-confiuration in OMNeT++:

\begin{lstlisting}[language=oppini,frame=none]
eventlogmanager-class = "PCAPNGEventlogManager"
\end{lstlisting}

The PCAPng manager uses the packet serialization feature of the INET framework. Thus it is able to write packets for all protocols and layers as long as a serializer was previously implemented and registered. This makes it applicable in other domains, e.g. wireless communication, or for verifying the implementation of models of protocols in OMNeT++.

\subsubsection{SQLite \& postgreSQL} The SQLite and postgreSQL ResultManagers allow to store simulation results into a SQLite database file or a postgreSQL database. Databases allow to perform complex queries on the simulation results. This can significantly speedup the process of obtaining network metrics, especially when huge parameter sets with various seeds were simulated.

\begin{figure}[!t]
  \centering
  \includegraphics[width=\linewidth]{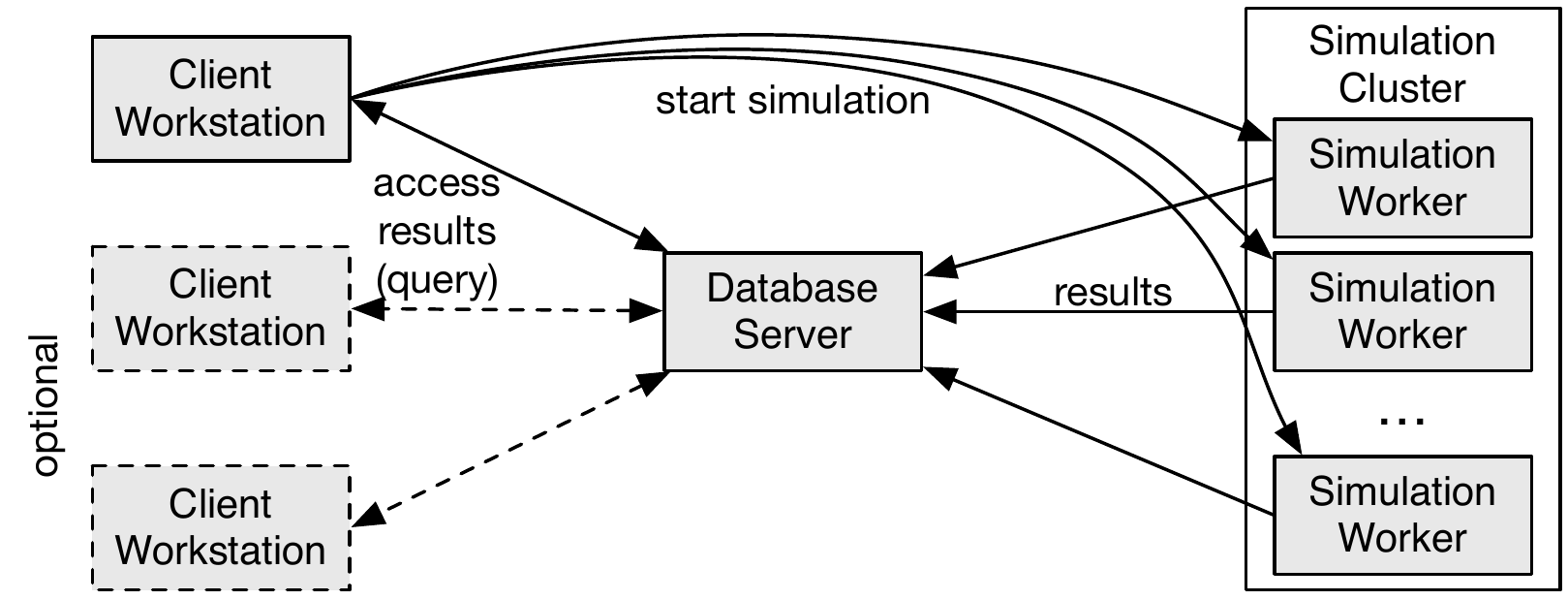}
    \vspace{-1.5em}
  \caption{Storing and accessing simulation results using a database (e.g. postgreSQL).}
  \label{fig:client_server}
   \vspace{-1em}
\end{figure}

\begin{figure*}[!b]
  \centering
   \vspace{-1em}
  \includegraphics[width=1\textwidth]{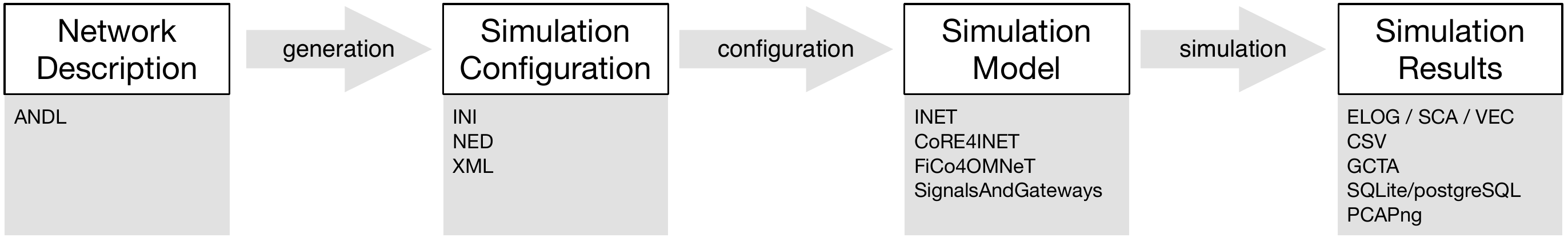}
  \vspace{-1em}
  \caption{Workflow of simulation projects -- From network description to result analysis}
  \label{fig:workflow}
\end{figure*}

SQLite is a file based database that is fast and efficient. The SQLite ResultManager is in most cases slightly slower than OMNeT++ vector and scalar managers, but produces smaller result files.
The SQLite database is always stored on the machine executing the simulation. As the database is locked when it is opened for writing (similar to a OMNeT++ result file) several simulation processes running concurrently cannot write to the same database. Thus different parameter sets or seeds simulated in parallel will have to use separate SQLite databases. A script provided with the result manager offers to merge several databases to enable queries containing results of several runs. As SQLite databases are regular files on the filesystem they can be easily transferred, archived, or read and manipulated using third party software.

The postgreSQL manager allows to write simulation results on a central database server in the network, while simulations are executed on a distributed cluster of nodes (see Figure \ref{fig:client_server}). Several users can access the results concurrently without the necessity to distribute the result files. This allows to transfer the load of the simulation as well as result analysis from the users workstations towards strong servers and large centralized storage systems. The drawback of this solution is a slight performance decrease due to the overhead of sending results over the network as well as delays due to the databases lock mechanisms when it is accessed concurrently. Using a database system, OMNeT++ simulations can be easily attached to a wide range of analysis tools, e.g. R using a database driver.

The database ResultManagers are enabled in the .ini configuration of the simmulation:
\begin{lstlisting}[language=oppini,breaklines=true,frame=none]
outputscalarmanager-class="cPostgreSQLOutputScalarManager"
outputvectormanager-class="cPostgreSQLOutputVectorManager"
postgresqloutputmanager-connection="dbname=testdb user=testuser password=testuser port=15432"
\end{lstlisting}

\subsubsection{GCTA}\label{sec:tlog} Writes out the previously introduced .tlog files for the GCTA plugin (see section \ref{sec:gcta}). The .tlog files contain information about the simulated topology as well as timing of cyclic messages. Traffic flows are aggregated based on their traffic classes. For example for messages using AS6802 the messages are grouped using the virtual link id, correspondingly for Ethernet AVB the stream id is used.

\subsubsection{Constraint Check} The ResultManager for constraint checks allows to define rules for output vectors. The defined bounds are not to be violated by the simulation. The ResultManager writes out a report of possible violations and can also end the simulation if a violation was detected. This way a parameter set with undesired results won't unnecessarily utilize CPU resources. For example, if a large number of simulations with different parameters and seeds is being batch executed, a run that does not comply with the requirements is immediately stopped when the violation occurs. This can significantly reduce the time required to simulate large sets of parameters.

Currently constraints include minimum and maximum checks, minimum and maximum checks using an average over a number of samples, over a time interval and checks using the sum of a vector. Constraints are configured in a XML format (see listing \ref{lst:constraints}) and can be easily extended.

\begin{lstlisting}[language=XMLconstraint,frame=lines,caption=Example of XML definition for the constraint check ResultManager, label=lst:constraints]
<constraints>
	<constraint module="Network.node1"
	   name="rxMessageAge:vector">
		<min>1.5</min>
		<max>1.7</max>
	</constraint>
	<constraint module="(.*)\.node2"
	  moduleIsRegex="true"
	  name="(rx|tx)MessageAge:vector"
	  nameIsRegex="true">
		<avg_min samples="10">1.5</avg_min>
		<avg_max samples="10">1.7</avg_max>
	</constraint>
</constraints>
\end{lstlisting}

\subsubsection{Multiple} This ResultManager allows to use several managers in parallel, enabling the user to write e.g. the legacy vector files in parallel with one of the new formats previously presented.

%
\section{Case Study}
\label{sec:case_study}
The simulation environment is a valuable and established part of our daily research and development.

\subsection{Simulation Workflow}
Our workflow (see figure \ref{fig:workflow}) starts with the network design. The ANDL is used to describe the required nodes, as well as the desired network topology, and the mapping of messages to different traffic classes. Afterwards our toolchain automatically generates an executable simulation configuration that is run using the simulation models for real-time Ethernet and fieldbusses. After the simulation run, the results are analyzed with the various result analyzers that are built into the OMNeT++ IDE, provided as additional plugins (e.g. the GCTA), or interconnected using databases and specialized output formats such as PCAPng.

\subsection{Simulating the Ethernet Backbone of a Prototype Car}
The presented simulation environment is used in several of our publications and is a base for the evaluation of new architecture concepts for in-car networks. So far, our largest project is the simulation of network designs for a real-world prototype car \cite{smkr-reicb-14}.

\begin{figure}[!t]
  \centering
  \includegraphics[width=\linewidth]{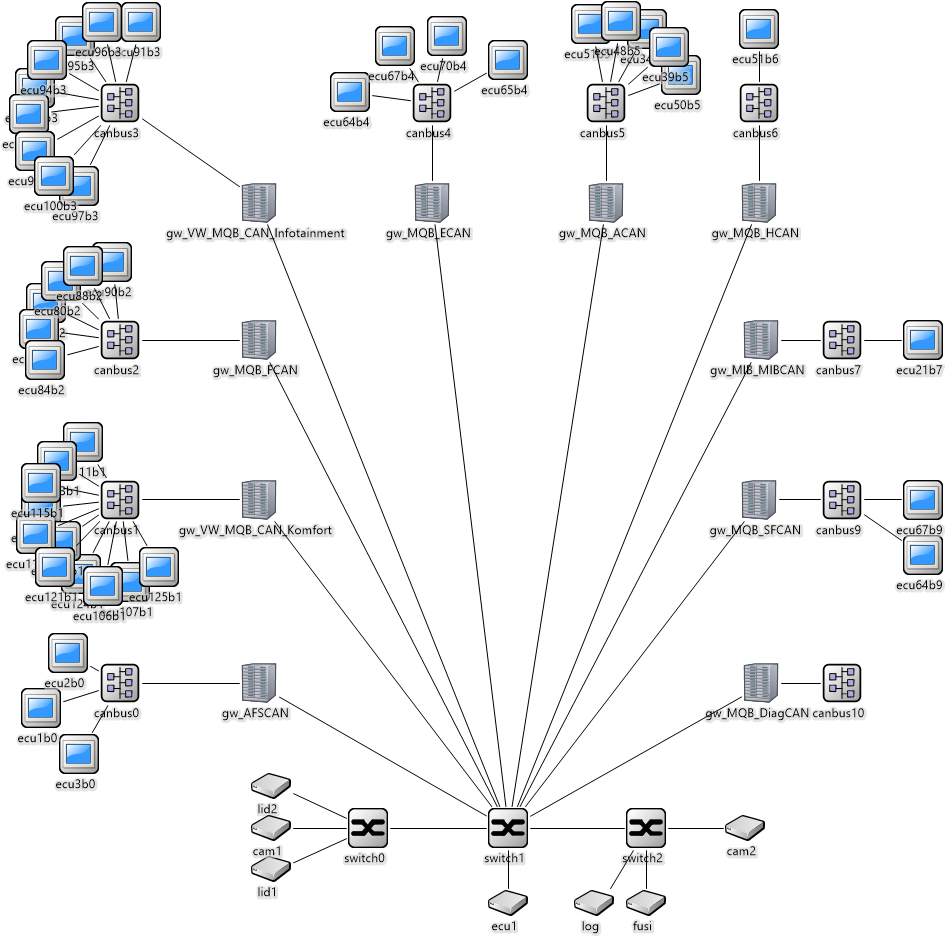}
  \vspace{-2em}
  \caption{Simulation of the Architecture of the RECBAR prototype}
  \label{fig:prototype}
  \vspace{-1em}
\end{figure}

The simulated prototype is a Volkswagen Golf 7 that was equipped with a real-time Ethernet backbone for the RECBAR research project. The car originally contains seven domain specific CAN busses interconnected over a central gateway node. For the prototype a real-time Ethernet backbone using three real-time switches and several additional nodes with high bandwidth applications such as high definition cameras and laser scanners were added. We simulate the prototype with real traffic patterns of the series car. Figure \ref{fig:prototype} shows the network.

In the simulation we are able to compare the results of the legacy network containg a central gateway node, with the results from the real-time Ethernet backbone. The results show that the real-time Ethernet solution can provide comparable end-to-end latency and jitter, while providing significant bandwidth reserves. The simulation allows us to evaluate the influences of different gateway aggregation strategies (see section \ref{sec:signalsAndGateways}) and additional best-effort background cross-traffic.

Compared to empirical measurements in the real-world prototype, the simulation allows faster assessment of a wide range of protocols and configuration parameters. Further, the transparent nature of the network simulation allows us to debug configuration errors much faster than using the real system. While real hard- and software requires us to apply probes and adopt code to trace errors and measure timing, the simulation already provides thousands of measuring points.

\section{Conclusion \& Outlook}
\label{sec:conclusion_outlook}
In-vehicle communication technologies are about to change from today’s fieldbusses to switched real-time networks. With network simulation on system-level the process of evaluating real-time and application protocols, developing new shaping strategies, assessing architectures, and predicting hardware requirements can be supported. We contribute a simulation environment consisting of simulation models as well as development and analysis tools to face the challenges in this upcoming technology transition. Our experiences with the simulation of in-car networks for prototypes and complex heterogeneous network architectures for future cars underlines the value of a uniform system-level simulation environment.

Our experiences with the development of OMNeT++ plugins and ResultManagers specialized for tasks in our domain of in-car network research show that for the daily work in development and research projects it is worth analyzing whether specialized tools can support the simulation workflow. With its Eclipse based IDE, OMNeT++ is a solid foundation for the development of such tools.

In our future work we focus on adding new technologies to our simulation suite, such as Ethernet with frame preemption as currently discussed in IEEE 802.1Qbu or the implementation of CAN with flexible data rate (CAN FD). We further work on refining our result analysis tools.

\vspace{-0.15em}
\section*{Download}
\vspace{-0.25em}
All simulation models as well as the analysis tools presented in this work are published open-source and can be downloaded from our website at:
\begin{center}
\vspace{-0.5em}
\url{http://sim.core-rg.de}
\vspace{-0.5em}
\end{center}

We further provide an Eclipse update site to simplify the installation of the presented OMNeT++ plugins. 
\vspace{-0.15em}
\section*{Acknowledgements} 
\vspace{-0.25em}
This work was supported by the German Federal Ministry of Education and Research (BMBF) under the project RECBAR.
\bibliographystyle{IEEEtran}

\end{document}